\documentclass[pre, aps,twocolumn,floatfix,superscriptaddress]{revtex4-1}
\usepackage{amsmath,amssymb,graphicx,cases}
\usepackage{epsfig}
\usepackage{booktabs}
\usepackage{textcomp}
\usepackage{epstopdf}
\usepackage{graphicx}
\usepackage{float}
\usepackage{braket}
\usepackage[normalem]{ulem}
\usepackage[utf8]{inputenc}
\usepackage{enumitem} 

\usepackage[normalem]{ulem}
\usepackage{dsfont}
\newcommand{\stkout}[1]{\ifmmode\text{\sout{\ensuremath{#1}}}\else\sout{#1}\fi}

\usepackage{fixmath}
\newcommand{\vect}[1]{\mathbold {#1}} 

\usepackage{color}
\definecolor{Blue}{rgb}{0.00, 0.00, 1.00}
\definecolor{Red}{rgb}{1.00, 0.00, 0.00}
\definecolor{Green}{rgb}{0.00, 0.60, 0.00}

\newcommand{\nn}{\nonumber}
\newcommand{\be}{\begin{equation}}
\newcommand{\ee}{\end{equation}}
\newcommand{\bea}{\begin{eqnarray}}
\newcommand{\eea}{\end{eqnarray}}

\usepackage{multirow}
\usepackage{float}
\usepackage{caption, subcaption}

\usepackage[colorlinks=true, urlcolor=blue, anchorcolor=blue, citecolor=blue,filecolor=blue,linkcolor=blue,menucolor=blue]{hyperref}

\begin{document}
\title{Large deviations in statistics of the convex hull of passive and active particles: A theoretical study}

\author{Soheli Mukherjee}
\email{soheli.mukherjee2@gmail.com}

\affiliation{Department of Environmental Physics,
  Blaustein Institutes for Desert Research, Ben-Gurion University of
  the Negev, Sede Boqer Campus, 8499000, Israel}

\author{Naftali R. Smith}
\email{naftalismith@gmail.com}

\affiliation{Department of Environmental Physics,
  Blaustein Institutes for Desert Research, Ben-Gurion University of
  the Negev, Sede Boqer Campus, 8499000, Israel}


\begin{abstract}

We investigate analytically the distribution tails of the area $A$ and perimeter $L$ of a convex hull for different types of planar random walks. 
For $N$ noninteracting Brownian motions of duration $T$ we find that the large-$L$ and $A$ tails behave as
$\mathcal{P}\left(L\right)\sim e^{-b_{N}L^{2}/DT}$
and
$\mathcal{P}\left(A\right)\sim e^{-c_{N}A/DT}$,
while the small-$L$ and $A$ tails behave as
$\mathcal{P}\left(L\right)\sim e^{-d_{N}DT/L^{2}}$
and
$\mathcal{P}\left(A\right)\sim e^{-e_{N}DT/A}$,
where $D$ is the diffusion coefficient.
We calculated all of the coefficients ($b_N, c_N, d_N, e_N$) exactly.
Strikingly, we find that $b_N$ and $c_N$
are independent of $N$, for $N\geq 3$ and $N \geq 4$, respectively.
We find that the large-$L$ ($A$) tails are dominated by a single, most probable realization that attains the desired $L$ ($A$).
The left tails are dominated by the survival probability of the particles inside a circle of appropriate size.
For active particles  and at long times, we find that large-$L$ and $A$ tails are given by
$\mathcal{P}\left(L\right)\sim e^{-T\Psi_{N}^{\text{per}}\left(L/T\right)}$
and
$\mathcal{P}\left(A\right)\sim e^{-T\Psi_{N}^{\text{area}}\left(\sqrt{A}/T\right)}$
respectively. We calculate the rate functions $\Psi_N$ exactly and find that they exhibit multiple singularities. We interpret these as DPTs of first order.
We extended several of these results to dimensions $d>2$.
Our analytic predictions display excellent agreement with existing results that were obtained from extensive numerical simulations.

\end{abstract}

\maketitle

\section{Introduction}

\subsection{Background}


Brownian motion (BM) is a fundamental stochastic process that appears in many systems ranging from biology, physics, finance, computer science and many more \cite{Majumdar2005}. BM represents a broad universality class in the sense that for many models of random walks (RWs) and/or models of active particles
-- particles that generate dissipative, persistent motion by extracting energy
from their surroundings \cite{Schweitzer2003, RBELS2012, MJRLPRS2013, FGGVWW2015, GKKRST23} -- the long-time typical behavior converges to that of a BM (passive).
Examples of active matter arise in many biological systems like cellular tissue behaviour \cite{TWA2009}, bacterial motion \cite{BB1972, Alt1980}, formation of fish schools \cite{VCJCS1995} and many more. Active particles exhibit a wide range of interesting behaviour like non-Boltzmann stationary state \cite{TC2008, DKMSS2019}, clustering at boundaries \cite{CT2015}, jamming \cite{SEB2016} etc. 
Different models of RWs and/or active particles are used to model various realistic systems such as movement of animals, self propelled particles, polymers etc.

 The convex hull of a trajectory is the minimal convex set that contains all of the points along the trajectory. Convex hulls of stochastic trajectories have attracted much recent interest and they find many applications. For instance, they provide a natural way to define the home range of animals, which is the  territory that the animal covers during a certain period of time.  The area and perimeter of the convex hull thus gives a quantitative measurement of the home range \cite{Worton1995, GPH1995}. Apart from the home range, the convex hull is a useful tool for analyzing other phenomena, for instance to detect different phases in intermittent stochastic trajectories \cite{LG2017}, or to study the spread of animal epidemics \cite{DMRZ2013}.

The mean perimeter and area of convex hull of a RW as a function of number of steps $t$ is known (in the limit of
large number of steps where it converges to BM) \cite{Letac1992}. In the large $t$ limit, the mean perimeter and area scale as $\langle L \rangle \sim \sqrt{t}$ and  $\langle A \rangle \sim t$ respectively. These results have been extended to many systems in the literature of physics  \cite{FMC2009, MCF2010, Claussen2015, WX15, Dewenter2016, Schawe2017, Schawe2018, Schawe2019, Schawe2020, MajumdarMori2021, SKMS22} and mathematics \cite{Eldan2014, KZ2016, VZ2018}. For example: Using the connection to extreme value statistics, exact results have been obtained for the mean perimeter and mean area for the convex hull of $N$ noninteracting  planar BMs  \cite{FMC2009, MCF2010}. In addition, the mean volume and surface area of the convex hull in arbitrary dimensions $d$ has been calculated for a single BM and Brownian bridge \cite{Eldan2014, KZ2016, VZ2018}, Le\'{v}y processes \cite{KLM2012}, and a single BM in a confined geometry \cite{BBMS2022}.

 It is natural to try to give a more detailed characterization of the convex hull statistics beyond the average behaviors.
Analytical calculations of the variance and  higher order moments of the distribution of these quantities are difficult \cite{SS1993}.  To our knowledge, the only case for which an analytic result exists for the higher moments is for Brownian bridges \cite{Goldman1996}, which is a BM constrained to end at its starting point. It is also called a "closed" BM (in contrast with the unconstrained "open" BM).

It is especially interesting and challenging to analytically understand the 
full distributions (including the large-deviation tails) of observables related to the convex hull (e.g., its perimeter or area). At present, they are rather poorly understood from an analytic point of view. One reason for this is that the convex hull constitutes a rather complex physical object, in the sense that it is affected, in a very nontrivial way, by temporal correlations of the physical process. Fluctuations of the convex hull perimeter or area therefore depend on the nonequilibrium dynamics of the stochastic process.
Moreover, these observables do not fall into one of the ``standard" existing categories of observables (e.g., ``dynamical observables"  \cite{Touchette2018}) for which the there exist generic theoretical frameworks for the study of large-deviation statistics.


In contrast, extensive numerical studies of the properties of convex hulls have been performed.
These studies investigated not only the regime of typical fluctuations, but also used advanced importance-sampling techniques to probe far into the the large-deviation regimes, describing convex hulls that are much larger or much smaller than average. The full distributions of the observables like area ($A$) and perimeter ($L$)  in $d=2$ dimensions, as well as their extensions to higher dimensions volume ($V$) and surface area  ($\mathcal{A}$) have been numerically computed for different types of stochastic processes.  These include the single planar BM or bridge \cite{Claussen2015}, multiple BMs \cite{Dewenter2016}, BM in higher dimensions \cite{Schawe2017}, self avoiding RW \cite{Schawe2018, Schawe2019}, run and tumble particle (RTP) \cite{Schawe2020},   BM with resetting \cite{MajumdarMori2021}. These numerical simulations were able to calculate the atypical fluctuations with probabilities that are in some cases extremely small, of order $10^{-1000}$ or even less. However, a comprehensive theory describing these numerical results is still lacking. Analytically there has been partial recent progress in mathematical literature for the `right' tails for the perimeter and area of a single  random walker (under certain assumptions) \cite{Visotsky2021, Visotsky2023}.

In this paper, we calculate exactly the distribution of the tails of $A$, $L$, $V$ and $\mathcal{A}$ for different types of stochastic processes  by studying  the large deviation function (LDF) \cite{hugo2009, Touchette2018} encoding the dynamics. One of the important features of these LDFs that they may exhibit singularities which can be interpreted as  dynamical phase transitions (DPTs) \cite{exclusion, glass, kafri, exclusion1, singularities, baek, baek1, NemotoEtAl19}. We consider both passive and active particles.  We find that the physical picture in the right and left tails of the distributions is markedly different. As shown below, the right tails are dominated by a single, optimal large-scale trajectory of the process. In contrast, the left tails are dominated by realizations of the process that remain within a circle of appropriate size. The understanding of these physical pictures is what enables us to obtain the distribution tails analytically.

 \subsection{Model definitions}

 Let us now define precisely the theoretical models for which we aim to study the convex-hull area and perimeter distributions.

 \begin{itemize}[wide, labelwidth=!, labelindent=0pt]
 
 \item  \textbf{ Brownian motion.}
 The motion of a BM in  arbitrary dimension is described by the following Langevin equation:
\be
\label{Langevin1}
\dot{\vect{r}}(t)=\sqrt{2 D}\; \vect{\xi}(t)
\ee 
where $\vect{r}(t)$ is the position of the  particle at time $t$ and $0 < t < T$, $D$ is the diffusion constant and $\vect{\xi}(t)$ are Gaussian white noises with $\langle \vect{\xi} (t) \rangle =0$ and $\langle \xi_i(t) \xi_j (t') \rangle = \delta_{ij} \delta(t-t')$. 
Here $\langle . \rangle$ denotes the ensemble average over realizations of the noise.

 In $d=2$, dimensional analysis yields the following scaling relations of  the perimeter and the area distribution in the physical parameters (i.e $L$, $A$, $D$ and $T$) \cite{Claussen2015}
\bea
\label{ScalingL}
\mathcal{P}(L) &=& \frac{1}{\sqrt{D T}} \,\,  P \Big(\frac{L}{\sqrt{D T}} \Big) \\
\label{ScalingA}
\mathcal{P}(A) &=& \frac{1}{D T} \,\,  P \Big(\frac{A}{D T} \Big) 
\eea
 where the functions $P$ and their arguments are dimensionless.

\begin{table*}
\begin{center}
\renewcommand{\arraystretch}{1.5}
\begin{tabular}{||c | c |c| c| c ||} 
 \hline
Dimensions & Types of walk & Observables & Tails & Open, Closed \\ [0.5ex] 
 \hline\hline
  d=2 & BM & Perimeter  & right & Open: $\mathcal{P}\left(L\right)\sim e^{- L^2 /16 DT}$;   Closed: $\mathcal{P}\left(L\right)\sim e^{- L^2 /4 DT}$\\ 
 \cline{4-5} 
 &  &  & left &  $\mathcal{P}\left(L\right)\sim e^{-4 \pi^2 x_1^2 DT / L^2}$;    \\ 
 \cline{3-3}  \cline{4-5} 
&  & Area  &  right & Open: $\mathcal{P}\left(A\right)\sim e^{- \pi A / 2 DT}$;   Closed: $\mathcal{P}\left(A\right)\sim e^{- \pi A /  DT}$   \\ 
 \cline{4-5} 
&   &  & left &  $\mathcal{P}\left(A\right)\sim e^{- \pi x_1^2 DT /  A}$;     \\ 
 \cline{2-5}
& $N \ge 3$ BMs & Perimeter  & right &$\mathcal{P}\left(L\right)\sim e^{-b_{N}L^2/DT},\quad b_{N}=1/36$. \\ 
 \cline{4-5} 
 &  &   & left &  $\mathcal{P}\left(L\right)\sim e^{-4 N \pi^2 x_1^2 DT / L^2}$   ;   \\ 
 \cline{3-3}  \cline{4-5} 
   & & Area  & right &  $\mathcal{P}\left(A\right)\sim e^{-c_{N}A/DT},\quad c_{3}=1/\sqrt{3}, \quad c_{N\ge4}=1/2$.  \\ 
 \cline{4-5} 
&   &  & left &  $\mathcal{P}\left(A\right)\sim e^{- N \pi^2 x_1^2 DT /  A}$;  \\ 
 \cline{2-5}
 & Active particles &  Perimeter  & right &  $\mathcal{P}\left(L\right)\sim e^{-T\Phi\left(L/2T\right)}$  \\ 

 \cline{3-3}  \cline{4-5} 
 &  & Area  & right & $\mathcal{P} \left(A\right)\sim e^{-T\Phi\left(\sqrt{2\pi A}/T\right)}$  \\ 
 
 \cline{2-5}
& $N \geq 3$ active particles &  Perimeter  & right & $\mathcal{P}\left(L\right)\sim e^{-T\min\limits_{M}\Phi(\alpha_{M}L/T)},\quad\alpha_{M}=\left(2M\sin{\frac{\pi}{M}}\right)^{-1}$. \\ 
 
 \cline{3-3}  \cline{4-5} 
  & & Area  & right &  $\mathcal{P}\left(A\right)\sim e^{-T\min\limits_{M}\Phi(\beta_{M}\sqrt{A}/T)},\quad\beta_{M}=\left(\frac{M}{2}\sin{\frac{2\pi}{M}}\right)^{-1/2}$.  \\ 
 
 \hline \hline
 $d>2$ & BM & Surface area  & right &  $\mathcal{P}\left(\mathcal{A}\right)\sim e^{-l_{\mathcal{A},d}\mathcal{A}^{2/\left(d-1\right)}/DT}; \quad l_{\mathcal{A},3} = \pi/4$ \\ 
 \cline{4-5} 
 &  &  & left &  $\mathcal{P}\left(\mathcal{A}\right)\sim e^{-  \tilde{f}_{d} D T  (\tilde{\mathcal{A}_d}/\mathcal{A})^{2/\left(d-1\right)}}$  \\ 
 \cline{3-3}  \cline{4-5} 
&  & Volume  &  right &  $\mathcal{P}\left(V\right)\sim e^{-l_{V,d} V^{2/d}/DT}; \quad  l_{V,3}\simeq 5.3 $  \\ 
 \cline{4-5} 
&   &  & left &  $\mathcal{P}\left(V\right)\sim e^{-\tilde{f}_{d}DT\left(\tilde{V}_{d}/V\right)^{2/d}}$
\\
 
  \hline
\end{tabular}
\caption{
Behaviors of the left and right tails of the  distributions of the area $A$ and perimeter  $L$ of the convex hull for different planar 
 stochastic processes of duration $T$: BM  with diffusion coefficient $D$,  $N$ (multiple) BMs, active particles and  $N$ 
active particles,  and their extensions to $d>2$ dimensions  (i.e., the volume $V$ and surface area $\mathcal{A}$). Here $x_1 = 2.4048\dots$ is the first zero of the Bessel's function $J_0(x)$. The case $N=2$ is very simply related to the case $N=1$ and the relation is given by Eq. \eqref{N=2}. $\Phi(\vect{z})$ is the rate function that describes the distribution of the position of the active particle at long times, see Eq.~\eqref{pos_dist_active}. For the RTP,  $\Phi(\vect{z})$ is given in Eq. \eqref{actionRTP}. The $\alpha_M$ and $\beta_M$ are the coefficients of the rate function  $\Phi(\vect{z})$ for the $N \geq 3$ active particles given by Eq. \eqref{alphabeta}. 
 $\tilde{f}_{d}$ is the smallest eigenvalue of the minus Laplace operator on the $d$-dimensional ball of unit radius with absorbing boundary conditions, 
and $\tilde{V}_d$ and $\tilde{A}_d$ are the volume and surface area, respectively, of the ball of unit radius.
Trajectories are open unless stated otherwise.  
}
    \label{tab:different rws}
\end{center}
\end{table*}

\item  \textbf{Active particles.}
A generic theoretical model for active particles can be written as
\be \label{Langevin_activeparticle}
\dot{\vect{r}}(t)= \vect{\Sigma} (t)
\ee
where $\vect{\Sigma}(t)$ represents a noise term that
originates in the self-propulsion of the particle  with correlation time  $1/ \tau$.  While most of our investigations for active particles will be quite general (under fairly mild assumptions as detailed below), for the sake of concreteness we will also give explicit results for particular models.
Two of the most extensively studied models  in $d=2$ are: Active Brownian particle (ABP) \cite{BLLRVV2016} with the following equation of motion
\be
\label{langeviv_abp}
\dot{x}=v_0 \cos{\theta}(t),\quad
 \dot{y}=v_0 \sin{\theta}(t), \quad \dot{\theta} = \sqrt{2 D_r} \, \eta(t)
\ee
where $\eta(t)$ is a Gaussian white noise with 
$\left\langle \eta\left(t\right)\right\rangle =0$
and $\left\langle \eta\left(t\right)\eta\left(t'\right)\right\rangle =\delta
\left(t-t'\right)$, $v_0$ is the constant speed of the particle and $D_r$ is the rotational diffusion constant.

Another one is RTP \cite{TC2008} moving on the two-dimensional $x-y$ plane. 
\be \label{langevin_rtp}
\dot{\vect{r}}(t)= v_0 \,  \vect{\sigma} (t)
\ee 
 where again $v_0$ is the constant speed of the particle and $\vect{\sigma} (t)$ is the coloured noise. It is unit vector that reorients at some constant rate $\gamma$  to a new orientation that is randomly chosen uniformly from the unit circle.  Both of these models converge to BM in appropriate limits. For instance, for the RTP,
 in the limit $\gamma \to \infty$ and $v_0 \to \infty$, keeping the ratio $\frac{v_0^2}{2 \gamma} = D_{\text{eff}}$ fixed (where $D_{\text{eff}}$ is the effective diffusion coefficient), $\vect{\sigma}(t)$ becomes white noise and the typical fluctuations of the active models reduce to BM with diffusion coefficient $D=D_{\text{eff}}$ \cite{SBS2020}.

 \item  \textbf{Random walks.}
One can also consider random walks that are discrete in time and/or in space \cite{Letac1992, MCF2010, Claussen2015, WX15, Dewenter2016, Schawe2017, Schawe2018, Schawe2019}. As we will show below, our mathematical formalism that deals with active particles also addresses such random walks, under certain assumptions regarding the distribution of step sizes and durations.

 \end{itemize}

 The rest of the paper is organized as follows. Our results are all concentrated in Sec.~\ref{sec:results}.  We begin in subsection \ref{sec:singleBM} by calculating the right and left tails for the area and perimeter distributions for a single planar BM. We extend these results to $N$ non-interacting BMsin subsection \ref{sec:multipleBMs}.
 In subsection \ref{sec:active}, we extend the results to active particles in $d=2$, uncovering a remarkable sequence of DPTs for the case of multiple RTPs.
 In subsection \ref{sec:higherDim} we consider higher dimensions, focusing on the volume and surface area distributions for a single BM in $d=3$.
 We conclude with a  discussion in Sec. \ref{sec:discussion}.  Several technical details are given in the Appendices.

\section{Results} 
\label{sec:results}

The simplest case is the single BM in $d=2$. We therefore begin with a full analysis of the two tails for this case, followed by extensions to multiple and/or active particles. Then we treat the case of a single BM in $d=3$.
The results are summarized in the table \ref{tab:different rws}.

\subsection{Single planar Brownian motion}
\label{sec:singleBM}

\subsubsection{Right tail: Large area (A) / large perimeter (L)}

Let us begin from the simplest case of a single BM. From the scaling forms \eqref{ScalingL} and \eqref{ScalingA} one immediately finds that large perimeters or areas are mathematically equivalent to the short-time and/or weak-noise limit,
i.e $DT \ll  A $ or $DT \ll  L^2 $. The probability is dominated by the most probable path constrained on a given value of the observable of interest ($A$ or $L$) in a short $T$ limit. In this short $T$ / weak noise limit, the optimal fluctuations method (OFM) \cite{Onsager, MSR, Freidlin, Dykman, Graham, Falkovich01, GF, EK04, Ikeda2015, Grafke15, MeersonSmith19, SmithMeerson19, Meerson2023} gives the equation of the optimal path of the motion constrained to a given value of the observable of interest. The path probability of the trajectory $\vect{r}(t)$ is given by
 \be \label{Prob_trajectory}
\mathcal{P}(\vect{r}(t)) \sim e^{-s/2 D}
\ee
where $s[\vect{r}(t)]$ is the Wiener action \cite{Majumdar2005}
\be 
\label{WienerAction}
s[\vect{r}(t)] = \frac{1}{2} \int_0^T \dot{\vect{r}}(t)^2 \, dt \, .
\ee
In the small-$DT$ limit, we apply the saddle-point approximation and thus find that the dominant contribution to $\mathcal{P}(A)$ ($\mathcal{P}(L)$) comes from the minimizer $\vect{r}(t)$ (the "optimal trajectory") of the action constrained on the value of the area (perimeter). 
The action is minimized by motion with constant speed $|\dot{\vect{r}}(t)|=$ constant with the additional constraints \cite{MeersonSmith19, SmithMeerson19, Meerson2023, VCB23} (see also Appendix \ref{OFM derivation}). The action of the optimal trajectory is thus given by
\be \label{OFM}
- \ln{\mathcal{P}} \simeq s = \frac{\mathcal{L}^2}{4 D T} 
\ee
where $\mathcal{L}$ is the length of the trajectory  and $\ln$ denotes the logarithm in the natural basis. The problem thus reduces to minimizing $\mathcal{L}$ constrained on a given value of the observable (area or perimeter).

\begin{figure}
     \centering
         \includegraphics[scale=0.5]{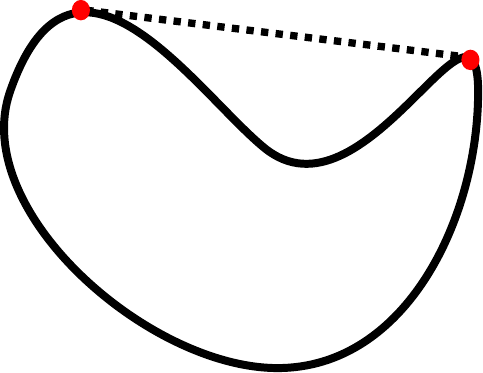}
         \caption{
         Schematic diagram of the convex hull of a trajectory of a BM. 
         To minimize the length of the trajectory $\mathcal{L}$ constrained on the convex hull area $A$ or perimeter $L$, it is more favourable to follow the dotted (straight) line than the solid line, when going between the two points marked in the figure.}
         \label{isoperimetric_projection}
     \end{figure}

 Let us start by studying the closed case which is a little simpler. Without loss of generality we consider only trajectories that are themselves the boundary of a convex shape, because  if this is not the case, one can always find a shorter trajectory with the same convex hull, see Fig.~\ref{isoperimetric_projection}. Under this assumption $\mathcal{L}$ equals the perimeter $L$ of the convex hull.

\begin{figure}
     \centering
         \includegraphics[scale=0.5]{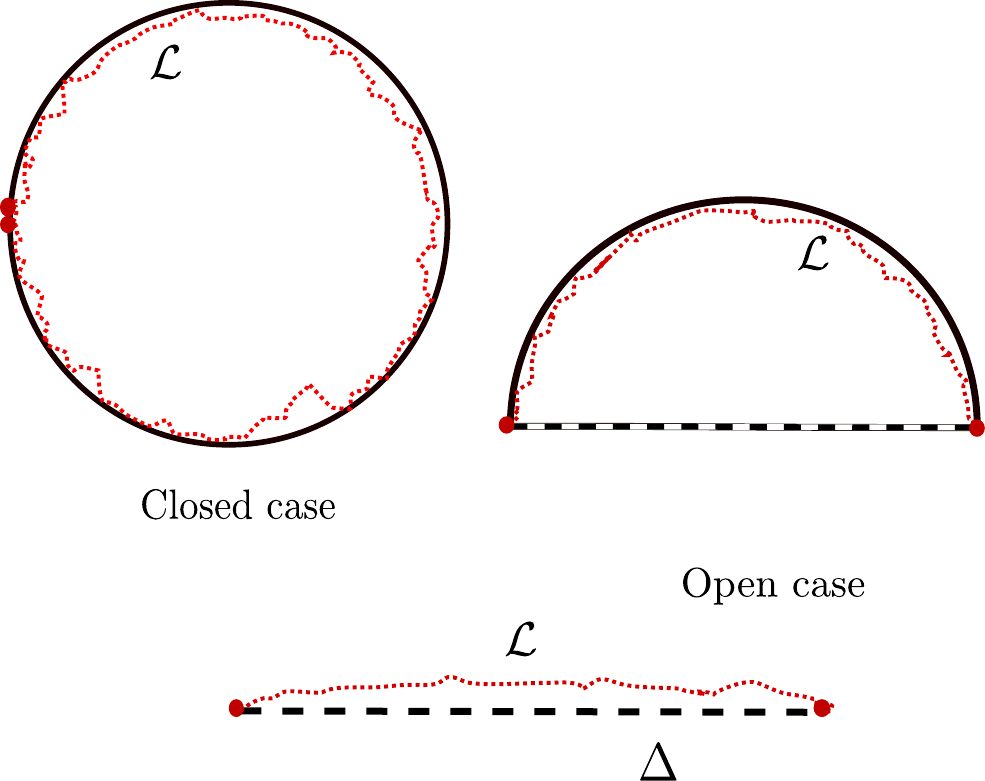}
         \caption{
          Solid lines: Trajectories of minimal length $\mathcal{L}$, corresponding to the right tails of the area and perimeter distributions. 
         For the area distribution, both the open and closed BMs are plotted.
         Schematic diagrams of the convex hulls formed by BMs of length $\mathcal{L}$, corresponding to the right tails of the area and perimeter distributions. 
          The dotted lines are a schematic of realistic realizations that attain large, but finite area (or perimeter).
         For the closed case, the hull is a circle for a fixed area $A$ and for open case, the hull is a half circle for a fixed $A$ and line segment for fixed $L$. }
         \label{convex hull schematic}
     \end{figure}

 Consider first the problem of minimizing $\mathcal{L}$ constrained on $A$. By the argument above, the problem reduces to that of finding a shape of minimal perimeter that encloses a given area -- the isoperimetric problem -- whose solution has been known for a very long time \cite{isoperimetric, isoperimetric2}: It is a circle of area $A$ (see Fig.~\ref{convex hull schematic} (a)). 
 The area and perimeter of a circle are related via $A = L^2/(4 \pi)$.
Hence the probability of the right tail behaves as [recalling that $\mathcal{L} = L$ and using Eq.~\eqref{OFM}]
\be \label{rightareaBM}
- \ln{\mathcal{P}(A \gg  D T)} = s = \pi \frac{A}{ D T} \, .
\ee

     \begin{figure*}
\centering
     \begin{subfigure}[b]{0.45\textwidth}
     \centering
         \includegraphics[width=1.1\textwidth]{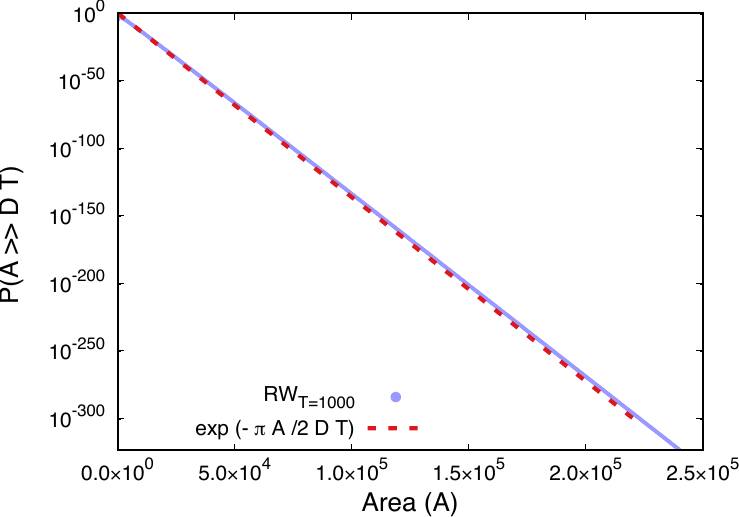}
          \caption{}
         \label{fig_area_op}
     \end{subfigure}
     \hfill
     \centering
     \begin{subfigure}[b]{0.45\textwidth}
     \centering
         \includegraphics[width=1.1\textwidth]{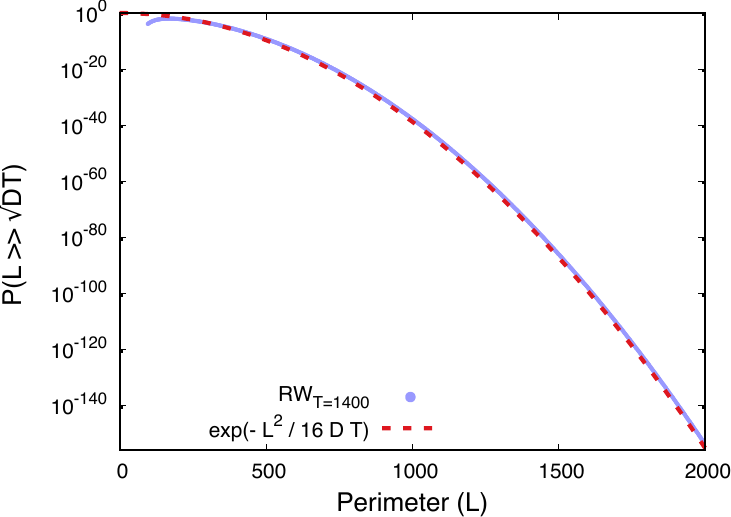}
         \caption{}
         \label{fig_peri_cl}
     \end{subfigure}
     \hfill
        \caption{ Right tails of the  distributions (in log scale) of the area (a) and perimeter (b) of the convex hull for an open planar BM. The blue  solid line depicts the data of $\mathcal{P}(A)$ from \cite{Claussen2015} and  can be seen to be in excellent agreement with the red dashed lines which denote  our theoretical predictions. Parameters are $D=1/2$, and $T=1000$ (a) and and $T=1400$ (b).}
        \label{fig_peri_area}
\end{figure*}

Let us now find the minimizer of $\mathcal{L}$ constrained on $A$ for an open BM. We assume that the particle begins at the origin (at time $t=0$) and without loss of generality, that it finishes on the $x$ axis at time $t=T$. We additionally assume that the trajectory is contained in the upper-half plane $y \ge 0$ (this assumption will be justified a posteriori). As for the closed case, the minimizer cannot have any concave sections.
The problem thus reduces to finding the curve of minimal length with the area under the curve constrained to a given value $A$.  This problem is known as Dido's problem \cite{isoperimetric, isoperimetric2}, which is the extension of the isoperimetric problem. The solution to Dido's problem is a semi-circle of area $A$ (see Fig.~\ref{convex hull schematic} (b)).
For the semi-circle, the length of the trajectory and the area are related via $A=\mathcal{L}^{2}/2\pi$.
Hence, the probability of the right tail for a open BM is given by
\be
\label{rightareaBMopen}
- \ln{\mathcal{P}(A \gg  D T)} = s = \frac{\pi}{2} \frac{A}{ D T} \, .
\ee
This optimal trajectory is in agreement with rigorous results from the mathematical literature which were obtained for general open RWs \cite{Visotsky2021}.
The result \eqref{rightareaBMopen} is also in excellent agreement with the numerical data of Ref.~\cite{Claussen2015}, see Fig.~\ref{fig_peri_area} (a).

Now let us solve the minimization problem of the length $\mathcal{L}$ constrained on the perimeter $L$. For the closed case, as shown above, $L = \mathcal{L}$  and therefore the minimization problem has a very large degeneracy of solutions (any closed trajectory which is the boundary of a convex shape is a minimizer).
For open motion, we assume again that the trajectory stays in the upper-half plane (taking the endpoint to be on the $x$ axis) and hence the relation between $L$ and $\mathcal{L}$ is  $L = \mathcal{L} + \Delta$,
where $\Delta$ is the distance between the endpoints of the BM. 
The minimal $\mathcal{L}$ is obtained when the curve approaches straight line (see Fig.~\ref{convex hull schematic} (c)), and then one has $\Delta \simeq \mathcal{L}$ so
\be
L = \mathcal{L} + \Delta \simeq 2\mathcal{L}
\ee
 Hence, the right tail of $\mathcal{P}(L)$ behaves as
\be
-\ln{\mathcal{P}(L\gg\sqrt{DT})}\sim\begin{cases}
\frac{L^{2}}{16DT}\,, & \text{Open},\\[2mm]
\frac{L^{2}}{4DT}\,, & \text{Closed}.
\end{cases}
\ee
The results are in excellent agreement with the numerical data of Ref.~\cite{Claussen2015}, see Fig.~\ref{fig_peri_area} (b).

\subsubsection{Left tail: Small area (A)/ small perimeter (L)}

The small $A$ (or small $L$) limit with constant $T$ is mathematically equivalent, according to Eq. \eqref{ScalingA}, to the long-$T$ limit at constant $A$ (or constant $L$).
In other words, the particle must survive inside the convex hull itself for an unusually long time. 
The least unlikely way for this to happen is if the convex hull takes the shape of a circle of area $A$ (or perimeter $L$).
We thus argue that $\mathcal{P}(A)$ (or $\mathcal{P}(L)$) is, in the leading order, given by the survival probability inside a circle of area $A$ (or perimeter $L$). 
In the leading order, it does not matter whether the BM is open or closed (the difference only affects a short time window in the trajectory, close to $t=T$).

\begin{figure}
     \centering
         \includegraphics[scale=0.4]{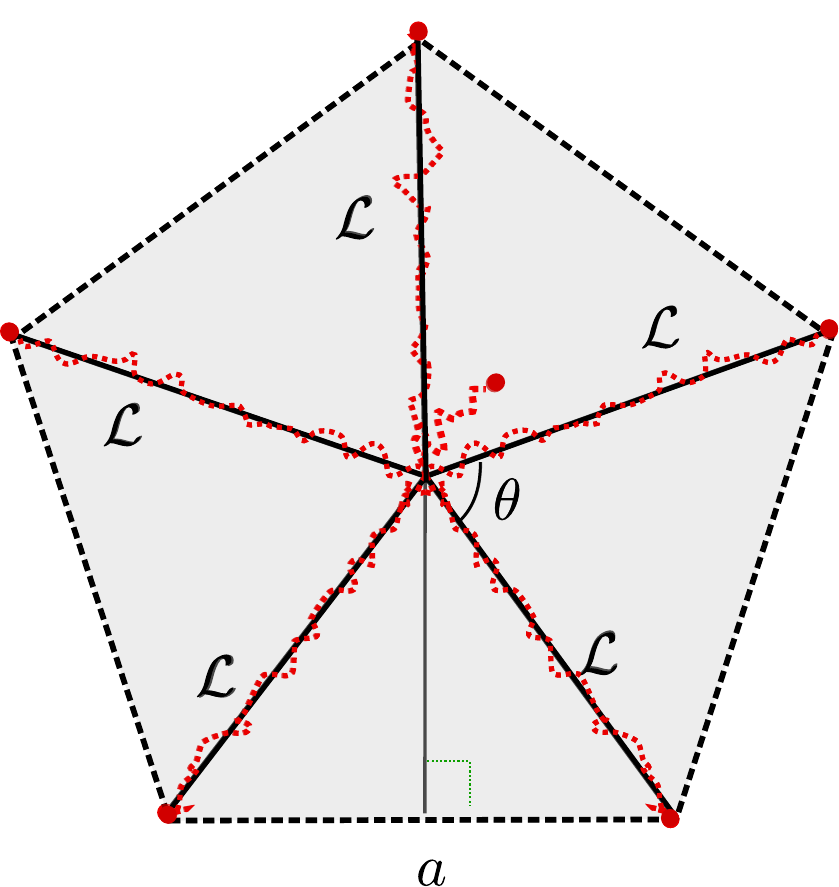}
         \caption{Schematic diagram of the convex hull formed by $N$ BMs of length $\mathcal{L}$. Here $N=6$, $M=5$, $\theta=\frac{2 \pi}{M}$ is the angle between two BMs participating in the solution.  As explained in the text, the solutions with $M=3$ and $M=4$ are the optimal solution for perimeter and area respectively.}
         \label{multiparticle hull}
     \end{figure}

In the long time limit (large $T$), the survival probability $S_{prob}\Big(T|R\Big)$ of a BM inside a $d$-dimensional ball of radius $R$ is, in the leading order, independent of the initial position within the ball, and is dominated by the smallest eigenvalue (in absolute value) of the Laplace operator with absorbing boundary conditions.
Therefore, it is given by (see e.g. \cite{TBA2016})
\be \label{surv_prob_lefttail}
-\ln{S_{prob}\Big(T|R\Big)}\simeq\begin{cases}
\frac{\pi^{2}}{4R^{2}}\,\,D\,T, & d=1,\\[2mm]
\frac{x_{1}^{2}}{R^{2}}\,\,D\,T, & d=2,\\[2mm]
\frac{\pi^{2}}{R^{2}}\,\,D\,T, & d=3
\end{cases}
\ee
where $x_1 = 2.4048\dots$ is the first positive root of the Bessel
function $J_0(z)$.






Using this, we obtain the left tails for the area and perimeter distributions by considering the survival probabilities in a circle of given area (or of given perimeter), yielding
\bea
-\ln {\mathcal{P}(A \ll  D T)} &\simeq& \pi x_1^2 \frac{ D T}{ A } \, ,\\[1mm]
-\ln{\mathcal{P}\left(L\ll\sqrt{DT}\right)} &\simeq& 4 \pi^2 x_1^2 \frac{ D T}{ L^2 } \, .
\eea
These formulas hold both for the open and closed BMs.
The scaling behaviors in these results are in agreement with the numerical results in \cite{Claussen2015} but the numerical coefficients observed there were different. We believe that this discrepancy is because the numerical simulations use a sufficient number of time steps to observe the continuous BM behavior.


\subsection{Multiple Brownian motions}
\label{sec:multipleBMs}

In this subsection, we consider $N>1$  non-interacting BMs, beginning from the case $N=2$ which is particularly simple. In this case, by concatenating the two trajectories we obtain a trajectory that could be considered as that of a single BM of twice the duration. Therefore, the distributions for $N=2$ are exactly related to those for $N=1$ via
\be \label{N=2}
\mathcal{P}\left(*,T\right)|_{N=2}=\mathcal{P}\left(*,2T\right)|_{N=1}
\ee
(where the time dependence is denoted explicitly, and $*$ represents area, perimeter or other such observables).
In particular, the distribution tails can thus be obtained immediately from the results reported above for the $N=1$ case.

Therefore, in the rest of this subsection we will assume that $N \ge 3$.
Let us consider first the right tail. Here,  following a similar approach to the one we used above for the case $N=1$, we find that this tail is dominated by the (multi-particle) trajectory that minimizes the sum of the Wiener actions $s_{\text{tot}}$, constrained on a given value of the area or perimeter. The minimum is obtained for trajectories for which each particle's speed is constant, and the sum of the Wiener actions is given by 
$s_{\text{tot}} = \left(\sum_{i=1}^{N}\mathcal{L}_{i}^{2}\right)/4DT$
where $\mathcal{L}_{i}$ is the length of the trajectory of particle $i$.

In order to proceed to solve this minimization problem we next assume that the full trajectories of the $N$ particles, except perhaps some of the endpoints, are all in the interior of the convex hull (this assumption will be justified a posteriori).
Thus only the endpoints of these trajectories are important, and the optimal trajectories must be straight lines. This simplifies the minimization problem considerably, because now we are just minimizing a function of the endpoints (and not a functional of the entire trajectories).

 For any $3 \le M \le N$ there exists a solution to this minimization problem for which $M$ of the particles travel the same distance $\mathcal{L}_i = \mathcal{L}$ in straight lines, and leaving the origin at equally-spaced angles, while the remaining $N-M$ particles remain near the origin. Thus, the convex hull that is formed by these trajectories is a regular polygon of $M$ sides (see Fig.~\ref{multiparticle hull} for an example with $N=6$ and $M=5$ in shaded region). The action $s_M = s_{\text{tot}}$ for each of these solutions 
 can be expressed as

\begin{figure}
         \includegraphics[scale=0.7]{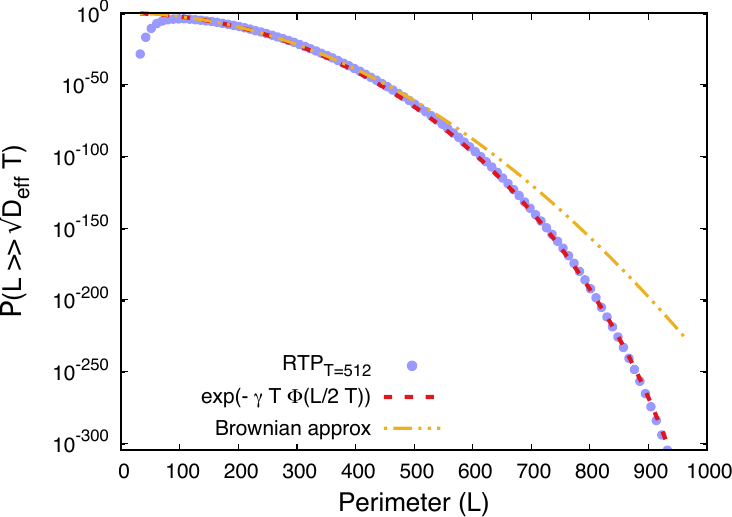}
         \caption{Right tail of the distribution $\mathcal{P}(L)$ of perimeter $L$  of the convex hull for a RTP  for walk length $T=512$ and  $\gamma=$ $v_0=1$. The blue circles depict the numerical data from \cite{Schawe2020} and the red dashed lines denote  our theoretical prediction. The yellow dot-dashed line denotes the  corresponding BM result, obtained by applying the parabolic Brownian approximation \eqref{BW_approx} of the rate function $\Phi(z)$. The Brownian approximation can be seen to correctly describe the near tail of the distribution, but it breaks down in the far tail.}
         \label{fig_peri_rtp_right}
     \end{figure}

\bea \label{action_multipleBW}
s_M(L) &=& \tilde{b}_{M} \,\frac{L^2}{ D T} \, ,  \\
 s_M(A)&=&  \tilde{c}_{M}\, \frac{A}{ D T} \, . 
\eea
The coefficients $\tilde{b}_{M}$ and $\tilde{c}_{M}$ are of geometric origin: They are calculated from the relation between $\mathcal{L}$ and $L$ or $A$ respectively for a regular $M$-sided polygon. For a $M$ sided polygon (for example Fig.~\ref{multiparticle hull} for $M=5$), the perimeter is $L= M a$ where $a = 2 \mathcal{L} \sin{\theta}$ is the length of each side of the polygon, which implies $L=\left(2M\sin{\frac{\pi}{M}}\right)\mathcal{L}$. Similarly, the area is given by $A=\left(M\sin{\frac{\pi}{M}}\cos{\frac{\pi}{M}}\right)\mathcal{L}^{2}$.  Finally, using that the total action is $s_{M}=M\mathcal{L}^{2}/4DT$, we obtain the coefficients
\be 
\tilde{b}_{M} = \frac{1}{16 M \sin^2{\frac{\pi}{M}}} \, , \qquad \tilde{c}_{M} = \frac{1}{2 \sin{\frac{2 \pi}{M}} } \, .
\ee

 These solutions all represent \emph{local} minima of the action $s_M$, and in order to find the \emph{global} minimum, we must now perform an additional minimization over $M = 3,4,\dots,N$.
This amounts to minimizing the coefficients $\tilde{b}_{M}$ and $\tilde{c}_{M}$ over $M$ for the perimeter and area, respectively.
These coefficients are tabulated in Table \ref{tab:multiple rws}. We find, and show explicitly in Appendix \ref{bmcm_plots}.1, that $M=3$ ($M=4$) is minimal (over all values $M=3,4,\dots$) for the perimeter (area) case. Therefore the optimal convex hull shape is an equalateral triangle whose center is the origin for the perimeter case and for the area case with $N=3$, and a square centered at the origin for the area case with $N \ge 4$.
So the probability is now expressed for any $N \geq 3$ as (see Table \ref{tab:different rws})

\bea
\label{NBMsRightTailL}
&& \!\!\!\!\!\!\!\!\! \mathcal{P}\left(L\right)\sim e^{-b_{N}L^{2}/DT},\quad b_{N}= \! \min_{3\le M\le N}\tilde{b}_{M}=\tilde{b}_{3}=\frac{1}{36} , \\
\label{NBMsRightTailA}
&& \!\!\!\!\!\!\!\!\! \mathcal{P}\left(A\right)\sim e^{-c_{N}A/DT}, \nn\\
&&\qquad c_{N}=\min_{3\le M\le N}\tilde{c}_{M}=\begin{cases}
\tilde{c}_{3}=\frac{1}{\sqrt{3}}, & N=3,\\[2mm]
\tilde{c}_{4}=\frac{1}{2}, & N\ge4.
\end{cases}
\eea
 Thus, rather remarkably, the right tail of the perimeter (area) distribution becomes, in the leading order, independent of $N$ for $N\ge 3$ ($N\ge 4$).
These theoretical predictions exhibit excellent agreement with the numerical observations from Ref. \cite{Dewenter2016}, see Appendix \ref{bmcm_plots}.2.

The left tails of the distributions of $A$ and $L$ for $N$ noninteracting particles behave very similarly to the single particle case. They are dominated by the survival probability of $N$ particles inside a circle of  
 appropriate area (or perimeter). 
Since the particles are noninteracting, the latter probability is simply given by that of a single particle, raised to the power $N$.
Using the middle line of \eqref{surv_prob_lefttail} together with the relations between the radius of a circle and its perimeter and area, one finds
\bea
\mathcal{P}\left(A\right)&\sim& e^{-\pi x_{1}^{2}NDT/A} \, ,\\
\mathcal{P}\left(L\right)&\sim& e^{-4\pi^{2}x_{1}^{2}NDT/L^{2}} \, .
\eea

%
%
%
%
\begin{table}
\begin{center}
\renewcommand{\arraystretch}{1.6}
\begin{tabular}{||c | c | c ||} 
 \hline
 $M$  & $\tilde{b}_{M} =  (16 M \sin^2{\frac{\pi}{M}})^{-1}$   & $\tilde{c}_{M} =  (2 \sin{\frac{2 \pi}{M}} )^{-1}$ \\
 [0.5ex] 
 \hline\hline
  $3$ & $1 / 36 = 0.027\dots$   & $1 / \sqrt{3} = 0.57\dots$  \\
  \hline
   $4$ & $1 / 32 =0.031\dots$   & $1 / 2 =0.5$ \\
   \hline
    $5$ & $\left(5+\sqrt{5}\right)/200=0.036\dots $  & $\sqrt{2/\sqrt{5}+5}=0.525\dots$ \\
    \hline
     $6$ & $1 / 24=0.041\dots$   & $1 / \sqrt{3}=0.577\dots$ \\
 \hline
  $7$ & $ 0.047\dots $  & $ 0.639\dots$ \\
    \hline
    $M \gg 1$  & $ \simeq M / 16 \pi^2$  & $ \simeq M^2 / 4 \pi$ \\
    \hline
\end{tabular}
\caption{Coefficients of the action $s_M(L)$ and $s_M(A)$ respectively given in Eq. \eqref{action_multipleBW}  for $N$ non-interacting BMs.  the minimum is obtained at $M=3$ ($M=4$) for the perimeter (area) case.}
    \label{tab:multiple rws}
\end{center}
\end{table}

%
%

\subsection{Active particles}
\label{sec:active}

 As described above, for a broad class of models of active models, the long-time typical behavior is diffusive, with an effective diffusion coefficient that can be found  (see examples above for the cases of the RTP and ABP). Thus, we expect that both the typical fluctuations and
the near tails of the area and perimeter distributions behave, at long times $T \gg \tau$  (where, to remind the reader, $\tau$ is the correlation time of the active noise), coincide with those of BM.
The signatures of activity are  expected to be found in the far tails of the distribution. In this subsection, we will focus on the behavior in the right tail. The left tail probabilities are still expected to be given by the long-time survival probabilities inside circles of appropriate sizes.  We do not attempt to calculate these survival probabilities in the current work, see however the recent Ref. \cite{TFC23} in which this was achieved for the ABP  (see also the related work \cite{IP23} where the RTP in $d=1$ was studied).

\subsubsection{Coarse graining}

For active particles, in the long time limit (where $T$ is much larger than the correlation time $\tau$ of the noise $\vect{\Sigma}$),
 one can coarse grain the noise $\vect{\Sigma}(t)$ that is coarse grained
 by averaging it over intermediate timescales $\tau\ll\Delta t\ll T$ \cite{HT09, Harris15, Jack19, JH20, AgranovBunin21, SmithFarago22, BTZ23, Smith2023, PLC24}.
\be \label{coarsefrained_noise}
\Bar{\vect{\Sigma}}(t)=\frac{1}{\Delta t}\int_{t}^{t+\Delta t}\vect{\Sigma}(t')\,dt' \, .
\ee
The probability of a coarse-grained noise history $\Bar{\vect{\Sigma}}(t)$ is given (in the leading order) by  the ``temporal additivity principle"
\be
\mathcal{P}[\Bar{\vect{\Sigma}}(t)] \sim \exp({- s[\Bar{\vect{\Sigma}}(t)]})
\ee
where the action $s[\Bar{\vect{\Sigma}}(t)]$ is now given by
\be \label{action_activeparticle}
s[\Bar{\vect{\Sigma}}(t)] = \int_0^T \Phi[\Bar{\vect{\Sigma}}(t)] \,\, dt \, .
\ee
Here we assume that the  long-time position distribution $\mathcal{P}\left(x,y,t\right)$ of the particle satisfies an large deviation principle (LDP) \cite{hugo2009, Touchette2018} with a rate function $\Phi(\vect{z})$, i.e., that
\be \label{pos_dist_active}
\mathcal{P}\left(x,y,t\right)\sim e^{-t\Phi\left(x/t,y/t\right)}\,.
\ee
$\Phi(\vect{z})$ is known for several standard models of active particles including RTP, ABP.
For the  RTP (Eq. \ref{langevin_rtp}) it was calculated exactly to be  \cite{DMS, SBS2020, Smith2023}
\be \label{actionRTP}
\Phi\left(\vect{v}\right)=2\gamma\phi\left(v/v_{0}\right),\quad\phi\left(z\right)=1-\sqrt{1-z^{2}} \, .
\ee
For ABP (see Eq. \ref{langeviv_abp}), the $\Phi$ was found exactly in \cite{PKS2016, KDAFPMB2018, BMRS2019} in terms of the smallest eigenvalue that gives periodic solutions to the Mathieu equation.
 In fact, the LDP \eqref{pos_dist_active} holds for many types of random walks in discrete time and/or space, where the rate function can be found from Cramér's theorem, see, e.g. \cite{hugo2009}.

The coarse-grained Langevin equation, obtained by  replacing $\vect{\Sigma}(t)$  by $\Bar{\vect{\Sigma}}$ in Eq.~\eqref{Langevin_activeparticle} is
\be
\label{LangevinCoarseGrained}
\dot{\vect{r}}(t)=   \Bar{\vect{\Sigma} }(t) \, .
\ee
For rotationally invariant statistics of the noise $\vect{\Sigma}(t)$, the $\Phi[\vect{z}]$ is also rotationally symmetric $\Phi[\vect{z}] = \Phi[z]$. 

As in the case of BM, the minimizer of \eqref{action_activeparticle} is obtained for trajectories for which the argument of $\Phi$ is of constant modulus which simply equals the speed $\mathcal{L}/T$ (see Appendix \ref{OFM derivation}). Therefore, we simply get 
\be
\label{sOfLengthActive}
s=T\Phi\left(\mathcal{L}/T \right)
\ee
where $\mathcal{L}$ is the length of the trajectory.  Thus, we find that here too, the problem boils down to minimizing $\mathcal{L}$ under the constraints.




\subsubsection{Right tail for a single active particle}

Using Eq.~\eqref{sOfLengthActive}, we find that the right tails of the area and perimeter distributions can be calculated by replacing  $\mathcal{L}^2 / 4DT$ in Eq. \eqref{OFM} by $T \Phi(\mathcal{L}/T)$. Thus, the right tails of the perimeter and area distributions  for the closed and open cases are given by:
\bea
\label{rightTailActiveL}
\ln{\mathcal{P}\left(L\right)}&\simeq&\begin{cases}
-\gamma T\Phi(L/T), & \text{Closed},\\[2mm]
-\gamma T\Phi(L/2T), & \text{Open},
\end{cases}\\[1mm]
\label{rightTailActiveA}
\ln{\mathcal{P}\left(A\right)}&\simeq&\begin{cases}
-\gamma T\Phi(\sqrt{4\pi A}/T), & \text{Closed},\\[2mm]
-\gamma T\Phi(\sqrt{\pi A}/T), & \text{Open}.
\end{cases}
\eea

 To give an explicit, concrete example, let us  consider the RTP.   For the RTP, using Eq.~\eqref{actionRTP} the right distribution tails  for the open case are given by
\bea \label{eqRTP_single}
\ln{\mathcal{P}(L)} \simeq - \gamma \Bigg (T - \sqrt{T^2 - \frac{L^2}{v_0^2}} \Bigg ); \quad L\gg \sqrt{D_{\text{eff}} T}\\
\ln{\mathcal{P}(A)} \simeq - \gamma \Bigg ( T - \sqrt{T^2 - \frac{4 \pi A}{v_0^2}} \Bigg ); \quad A\gg D_{\text{eff}} T 
\eea
where $D_{\text{eff}}$ is the effective diffusion coefficient. Fig.~\ref{fig_peri_rtp_right} shows the right tail of the perimeter for $T=512$, which fits well with the experimental data from \cite{Schawe2020}. 
For small $|z| \ll   1$, the $\Phi(z)$ can be approximated as a parabola
\be \label{BW_approx}
\Phi(z) \simeq \frac{z^2}{2}
\ee
which corresponds to passive limit and the action in Eq.~\eqref{action_activeparticle} reduces to the Wiener action \eqref{WienerAction} \cite{Majumdar2005}, which is the BM limit.  As a result, the near right tail of the distribution, $\sqrt{D_\text{eff} T} \ll L \ll v_0 T$, coincides with that of the BM result, see Fig.~\ref{fig_peri_rtp_right}.



\subsubsection{Dynamical phase transitions for multiple active particles}

 Let us now analyze the case of $N>1$ active particles.
We first consider the case $N=2$. 
 At $T \gg \tau$, the argument used above for BMs, to relate the cases $N=2$ and $N=1$, still holds (but only approximately). Thus, Eq. \eqref{N=2} still approximately holds at long times, both in the typical-fluctuations and large-deviation regimes.

     \begin{figure*}
\centering
     \begin{subfigure}[b]{0.47\textwidth}
     \centering
         \includegraphics[width=1.1\textwidth]{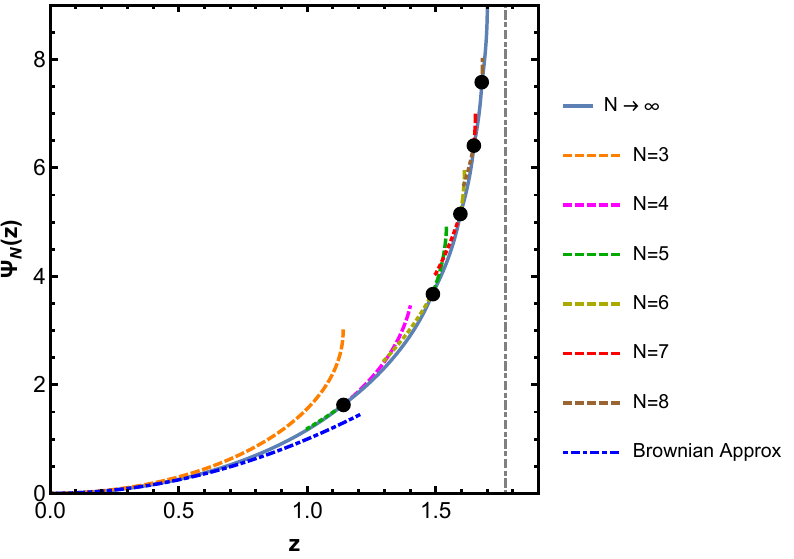}
         \caption{}
         \label{fig_area_multi_RTP_right}
     \end{subfigure}
     \hfill
     \centering
     \begin{subfigure}[b]{0.47\textwidth}
     \centering
         \includegraphics[width=1.1\textwidth]{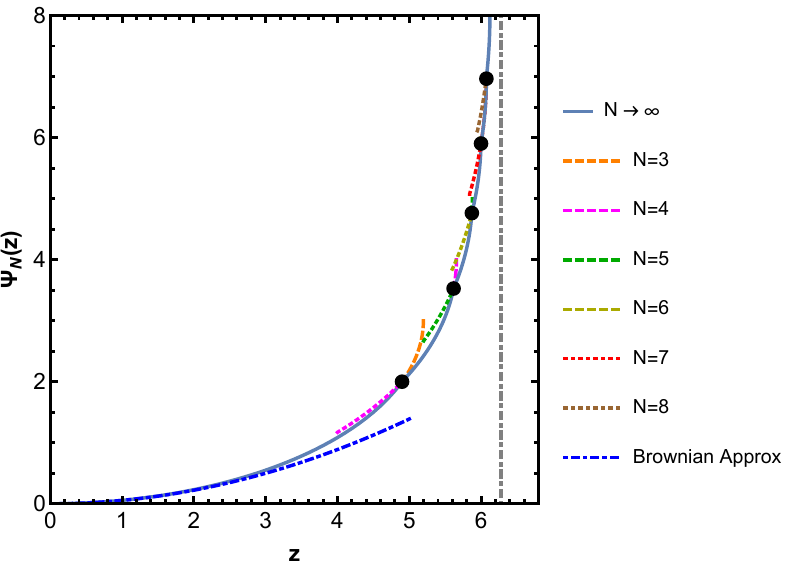}
         \caption{}
         \label{fig_peri_multi_RTP_right}
     \end{subfigure}
     \hfill
        \caption{
        The rate function $\psi_N(z)$   that describes  the right distribution tail for
        \textbf{(a)} Area $A$  and \textbf{(b)} Perimeter $L$ for $N$ RTPs. The solid blue line  depicts  $\psi_{N\to\infty}$  while the dashed lines correspond to $\psi_N$ for  finite values of   $N=3,4,\dots$ (from bottom to top). The dotted lines correspond to $\tilde{\psi}_M$ in regions where they are not optimal for any value of $N$  (again increasing with $M$ from bottom to top).  The circles denote the critical points $z_M$ at which the optimal solution changes from one local minima ($M$) to another local minima ($M+1$),  signalling a first-order DPT. 
         The blue  parabolic dot-dashed lines correspond to the BM approximations [for $N\ge 4$ in (a)], and are seen to give the correct asymptotic behavior at $z \ll 1$, describing the near right tail of the distribution.
        The  vertical, grey dot-dashed lines denote the  value that the area or perimeter cannot possibly exceed for any $N$, corresponding to a convex hull which is a circle of radius $v_0 T$.
        }
        \label{fig_multi_RTP}
\end{figure*}

 Let us now analyze the case $N \ge 3$,  focusing on the right tail.
  The analysis is similar to the case of $N\ge 3$  BMs studied above (see Sec. \ref{sec:multipleBMs}).
 One must minimize the sum of the coarse-grained actions
 $s_{\text{tot}} = T \sum_{i=1}^{N} \Phi(\mathcal{L}_{i}/T)$.
  We again assume that the trajectories of the $N$ active particles reside  within the interior of the convex hull, except for some of the trajectories' endpoints. Thus, the optimal paths are all straight lines, and the  problem reduces to that of minimizing $s_{\text{tot}}$ with respect to the endpoints of the trajectories. 
 Again we find that for each $3 \leq M \leq N$ there exists a solution in which $M$ particles exit the origin at equally-spaced angles, and each travel a distance of $\mathcal{L}_i = \mathcal{L}$, while the remaining $N-M$ particles stay near the origin, creating a convex hull whose shape is a regular polygon with $M$ sides, see Fig.~\ref{multiparticle hull}.
However, the actions $s_M= s_{\text{tot}}$ of these solutions,
  $s_M = MT \Phi(\mathcal{L}/T)$,
  are different to those of the BMs' case.  As a result, the optimal value of $M$  may too be different, and as we show below, it can in fact change within the tail,  leading to DPTs which are generically of the first order.

Therefore, 
the right tails of the perimeter and area distributions are determined by the solution with optimal $M$, i.e. that minimizes $s_M$,
\bea
\mathcal{P}\left(L\right)  &\sim&  \,\, e^{-T \Psi_N(  L /  T)} \, , 
\\
\mathcal{P}\left(A\right)  &\sim&  \,\, e^{-T \Psi_N(  \sqrt{A} /  T)} \, ,   
\eea
where the large-deviation functions $\Psi_{N}$ are related to the $s_M$'s above via 
\bea
\label{PsiNGeneralL}
\Psi_{N}\left(L/T\right)&=& \min_{3\le M\le N} M\Phi\left(\alpha_{M}L / T\right) \, ,\\ 
\label{PsiNGeneralA}
\Psi_{N}\left(\sqrt{A} / T\right)&=&\min_{3\le M\le N} M\Phi\left(\beta_{M}\sqrt{A} / T\right)\, .
\eea
Here the coefficients $\alpha_{M}$ and $\beta_{M}$ are calculated from the geometric relation between $\mathcal{L}$ and $L$ or $A$ respectively similar like the multiple BMs in Fig.~\ref{multiparticle hull} in Sec. \ref{sec:multipleBMs}, i.e $L=\left(2M\sin{\frac{\pi}{M}}\right)\mathcal{L}$ and $A=\left(M\sin{\frac{\pi}{M}}\cos{\frac{\pi}{M}}\right)\mathcal{L}^{2}$. Hence the $\alpha_{M}$ and $\beta_{M}$ are related to $\tilde{b}_M$ and $\tilde{c}_M$ as: 
\bea \label{alphabeta}
\alpha_{M} &=& 4 \sqrt{M \, \tilde{b}_M} = \frac{1}{2 M \sin{\frac{\pi}{M}}} \, , \\
 \beta_M &=&  2  \, \sqrt{\tilde{c}_M/ M} = \frac{1}{ \sqrt{ M \sin{\frac{\pi}{M}}  \cos{\frac{\pi}{M}}}}\, .
\eea

%

 Let us now give explicit results for the particular case of multiple RTPs. Here the rate functions are conveniently written as
\be  
\Psi_{N}\left(v\right)=2\gamma\psi_{N}\left(v/v_{0}\right)
\ee
where the $\psi_{N}$'s are dimensionless, and are given by 
\be
\psi_{N}\left(z\right)=\min_{3 \leq M \le N}\tilde{\psi}_{M}\left(z\right)
\ee
and $\tilde{\psi}_{M}$ is related to the dimensionless rate function $\phi$ in Eq.~\eqref{actionRTP} that describes the position distribution of a single particle,
\bea \label{phi_multipleRTP}
&&\tilde{\psi}_M\left(z\right)=M\phi\left(a_{M}z\right)=M\left(1-\sqrt{1-a_{M}^{2}z^{2}}\right)\nn\\
&&=\begin{cases}
M\left[1-\sqrt{1-\frac{z^{2}}{4M^{2}\sin^{2}{(\pi/M)}}}\,\right] & \text{(perimeter)}\\[3mm]
M\left[1-\sqrt{1-\frac{z^{2}}{M\sin{(\pi/M)\cos{(\pi/M)}}}}\,\right] & \text{(area)}
\end{cases} \nn\\
\eea
where $a_M= \alpha_M, \beta_M$ for perimeter and area respectively.

In the near right tails, the value of $M$ that dominates is the same as for the case of BMs.
For $T \gg \tau$, the right tail of the $N$ active particles behave like the $N$ noninteracting BMs where the action for $M=3$  in case of perimeter and $M=4$ in case of area dominates.
Further into the right tail (for sufficiently large $N$), however, activity dominates and the $M=4$ (for perimeter) and $M=5$ (for area) becomes the optimal solution. This leads to singularity in the rate function $\psi_N(z)$ which we interpret as a DPT, at a critical value of $z$, 
shown by the solid black circle in Fig.~\ref{fig_multi_RTP}. Further into the tail further successive DPTs occur from $M=i$ to $M=i+1$ for $i=3,...M-1$ in case of perimeter and for $i=4,...M-1$ in case of area. So the right tail of the distribution exhibits $N-3$ and $N-4$ DPTs for perimeter and area respectively. As the transitions occur when the graphs of two $\tilde{\psi}_M$'s cross each other, this transition is of first order in nature, i.e., the first derivative of $\psi_N$ jumps at the critical point.

    The coordinates of the critical points can be calculated by 
    requiring $\tilde{\psi}_M(z) = \tilde{\psi}_{M+1}(z)$, which, using Eq. \eqref{phi_multipleRTP} gives
  \be 
M\left(1-\sqrt{1-a_{M}^{2}z^{2}}\right) = (M+1)\left(1-\sqrt{1-a_{M+1}^{2}z^{2}}\right) \,.
  \ee
   The solution to this equation yields the critical points $z=z_M$, which are given by  
  \be
  z_M = - \frac{2 \sqrt{M (M+1)} \sqrt{a_{M+1}^2 (M+1)-a_M^2 M}}{a_M^2 M^2-a_{M+1}^2 (M+1)^2} \, .
  \ee
  The first few critical points are at 
  \be
  z_3 = 2\sqrt{6}=4.898\dots, \quad z_4 = 2 \sqrt{8 \sqrt{5}-10}=5.617\dots
  \ee
  for the perimeter, and at 
  \be
  z_4 = 1.141\dots, \quad z_5 = 1.489\dots
  \ee
  for the area respectively. The corresponding critical values of $L$ and $A$ are given by $L=z_{M}v_{0}T$ and $A=\left(z_{M}v_{0}T\right)^{2}$.

More generally, it is reasonable to expect that similar DPTs may occur in many models of active particle, and not just in the particular case of the RTP as shown here.
In particular, if the particle's speed is bounded e.g., for the ABP we conjecture that the qualitative picture is very similar to that of the RTP.



\subsection{$d>2$} \label{sec:higherDim}

\begin{figure}
    \centering
    \includegraphics[scale=0.6]{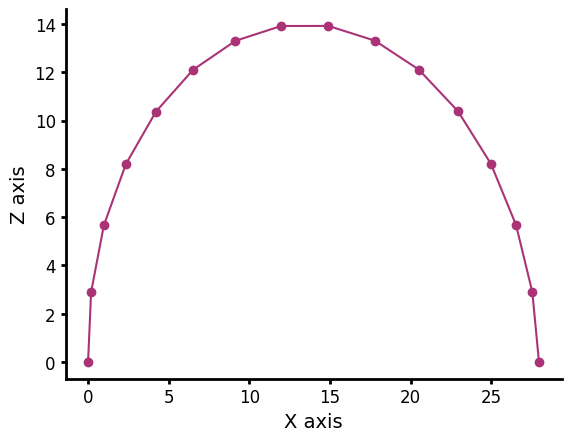}
    \caption{
    The shortest curve in the three dimensional space whose convex hull has a given surface area, obtained numerically using a gradient-descent algorithm.
    The figure shows the projection of the trajectory onto the  $xz$ plane.
     The $y$ components of all the points are  very small on  the scale of the figure.
}
    \label{optimal_surface_trajectory}
\end{figure}

Let us now consider the convex hull in higher dimensions ($d>2$).  The arguments above can be extended straightforwardly to study the volume ($V$) and surface area ($\mathcal{A}$) distributions of the convex hull in higher dimensions $d>2$.
For instance, for a single BM, this leads to the following behaviors in the tails:
\bea
\label{scalingGeneralD}
\mathcal{P}\left(V\right) &\sim&\begin{cases}
e^{-l_{V,d} V^{2/d}/DT}, & V\gg\left(DT\right)^{d/2}\\[1mm]
e^{-\tilde{f}_{d}DT\left(\tilde{V}_{d}/V\right)^{2/d}}, & V\ll\left(DT\right)^{d/2}
\end{cases}\\[1mm]
\mathcal{P}\left(\mathcal{A}\right)&\sim&\begin{cases}
e^{-l_{\mathcal{A},d}\mathcal{A}^{2/\left(d-1\right)}/DT}, & \mathcal{A}\gg\left(DT\right)^{\left(d-1\right)/2} \\[1mm]
e^{-  \tilde{f}_{d} D T  (\tilde{\mathcal{A}_d}/\mathcal{A})^{2/\left(d-1\right)}}, & \mathcal{A}\ll\left(DT\right)^{\left(d-1\right)/2}
\end{cases} \nn\\
\eea
Here $\tilde{f}_{d}$ is the smallest eigenvalue of the minus Laplace operator on the $d$-dimensional ball of unit radius  with absorbing boundary conditions, 
 $\tilde{V}_d$ and $\tilde{A}_d$ are the volume and surface area, respectively,
 of the ball of unit radius (see e.g. Ref. \cite{TBA2016}),
and $l_{V,d}$ ($l_{\mathcal{A},d}$) is a coefficient calculated from the geometric relations between the $\mathcal{L}$ and $V$ ($\mathcal{A}$) in $d>2$ dimensions.
%
The coefficients for the right tails are found by minimizing $\mathcal{L}$ constrained on $V$ (or $\mathcal{A}$). This minimization problem appears to become more difficult as $d$ is increased. The scaling behaviors \eqref{scalingGeneralD} were conjectured and numerically observed in Ref.~\cite{Schawe2017}.

 For $d=3$ we conjecture that the curve of minimal length constrained on a convex-hull surface area $\mathcal{A}$ is in fact confined to a plane ( which, without loss of generality, can be taken to be the $xz$ plane),
 and follows the edge a semi circle of area $\mathcal{A}/2$. The convex hull is thus a semi-circle shaped slice of infinitesimal width, so its surface area is $\mathcal{A}$.
The reasoning behind this conjecture is as follows.
 The curve is optimal with respect to deformations within the $xz$ plane -- this follows from the solution to the $d=2$ problem conditioned on the area $A$ (see above).
On the other hand, both the curve's length $\mathcal{L}$ and the convex hull's surface area $\mathcal{A}$ are mirror symmetric with respect to deformations of the curve in the $y$ direction (perpendicular to the curve). 
 It follows that for the semi-circular curve, the variation of $\mathcal{L}$ constrained on $\mathcal{A}$ indeed vanishes, and therefore this curve is a natural candidate for the (global) constrained minimizer.
 For this curve, $\mathcal{A}$ and $\mathcal{L}$ are related via
$\mathcal{A}/\mathcal{L}^2 = \pi$ 
and using this in Eq.~\eqref{OFM} leads to a coefficient of  $ l_{\mathcal{A},d=3}=\pi/4$.

 Very strong evidence in favor of our conjecture is that this coefficient is in agreement with  the numerically-observed coefficient from Ref.~\cite{Schawe2017}, see \cite{footnote:SchaweCoefficient}.  To further verify our conjecture, we numerically minimized $\mathcal{L}$ constrained on $\mathcal{A}$
using a gradient descent algorithm, and found that the shortest curve conditioned on a given surface area indeed appears to be a semi circle (see Fig.~\ref{optimal_surface_trajectory}). 
 To summarize, the scaling behaviors of the tails of the surface-area distribution in $d=3$ are given by
\be \label{d=3surface}
\mathcal{P}\left(\mathcal{A}\right)\sim\begin{cases}
e^{-\pi\mathcal{A}/4DT}, & \mathcal{A}\gg DT\\[1mm]
e^{-4\pi^{3}DT/\mathcal{A}}, & \mathcal{A}\ll DT
\end{cases}
\ee
 where we also plugged in the values of $\tilde{f}_{3}$ and $\tilde{A}_3$.

For the volume (in $d=3$) we were not able to obtain an analytical result for the coefficient $l_{V,3}$, but it was found numerically to be $\simeq 5.3$ in Ref.~\cite{Schawe2017}  (see \cite{footnote:SchaweCoefficient}).
The tail behaviors of the volume distribution are thus given by
\be \label{d=3volume}
\mathcal{P}\left(V\right)\sim\begin{cases}
e^{-5.3 V^{2/3}/DT}, & V\gg\left(DT\right)^{3/2}\\[1mm]
e^{- (4 \pi^2 / 3)^{2/3} DT/V^{2/3}}, & V\ll\left(DT\right)^{3/2}
\end{cases}
\ee
 where we plugged in the values of $\tilde{f}_{3}$ and $\tilde{V}_3$.

 One can extend the analysis to $N > 1$ particles and/or to active particles, as we did above for the case $d=2$, but we will not do so here. We do expect, however, that some of the qualitative features that we found in $d=2$ will hold in arbitrary $d$. In particular, we expect that:
(i) The relations found above between $N=2$ and $N=1$ hold in arbitrary $d$.
(ii) In the leading order, the right tails of $P(V)$ and $P(\mathcal{A})$ become independent of $N$ for sufficiently large $N$.
(iii) For $N$ active particles, the right tails of $P(V)$ and $P(\mathcal{A})$ will be described by rate functions that are simply-related to the rate function $\Phi$, and for sufficiently large $N$, first-order DPTs will occur.







\section{Discussion}
\label{sec:discussion}

 We analytically studied the tails of the distributions of area $A$ and perimeter $L$ of convex hulls formed by the motion of active or passive particles in the plane, and analogous quantities in $d>2$. We achieved this by identifying the scenario(s) that dominate the contribution to the probabilities of the rare events in question. Our findings are summarized in Table \ref{tab:different rws}.

In the left tails, the scenario is that of long-time survival of the particles inside a circle of appropriate size.
In the right tails, the OFM is valid, i.e., the probabilities are dominated by the most likely trajectory (or coarse-grained trajectory in the case of active particles) constrained on the observable.
Remarkably, we found that the right tails of $\mathcal{P}(L)$ and $\mathcal{P}(A)$ for $N$ BMs become, in the leading order, independent of $N$ at $N\ge 3$ and $N \ge 4$ respectively. This is because the optimal path involves significant motion of only three (four) of the particles for the perimeter (area) distribution.


 For a single arbitrary rotational-invariant active particle, we calculated the exact LDFs that describe the right tails of the distributions of the area and perimeter at times $T$ that are much longer than the microscopic characteristic timescale of the particle.  Remarkably, we found that these LDFs are simply related to the rate function $\Phi$ that describes the long-time position distribution of the active particle, see Eqs.~\eqref{rightTailActiveL} and \eqref{rightTailActiveA}.
We then extended these results to $N$ non-interacting active particles, see Eqs.~\eqref{PsiNGeneralL} and \eqref{PsiNGeneralA}. We found that, depending on $\Phi$, there may be DPTs in the right tails, signalling a sudden change in the behavior of the system as critical values of $L$ (or $A$) are crossed. We illustrated this by calculating the LDFs explictily for the case of the RTPs, where we found that for $N>3$ ($N>4$), the right tail of the perimeter (area) distribution exhibits $N-3$ ($N-4$) such DPTs, which are all of the first order. 
%
Note that although we stated our results in the context of active particles, they are more general and cover a broad class of models (e.g. RWs  that are discrete in time and/or space) for which the large-deviation principle \eqref{pos_dist_active} holds (with a rotationally-invariant $\Phi$).

 Finally, we have considered the distribution of surface area and hyper-volume of the convex hull for a BM in $d>2$ dimensions, with special emphasis on the case $d=3$. We calculated the left tails: They are again given by the survival probability inside a hyper-sphere of appropriate size. We were able to obtain the scaling behavior in the right tails up to a numerical constant, which is found by solving a minimization problem. This problem appears to be difficult to solve analytically in general, but we were able to  obtain its solution for the surface area case in $d=3$.
In other cases, on can solve the problem numerically.

 In many of the cases that we studied here, we were able to compare our theoretical predictions with existing numerical data, showing excellent agreement in the right tails in all cases. 
We hope that this work will stimulate additional numerical investigations (in particular, in the left tails where we believe that more extensive numerical work is needed in order to observe convergence to the theoretical results).

 Several interesting future directions of research remain. It would be interesting to  extend our results to non-rotationally-invariant active particles \cite{Visotsky2023},  and/or to the case in which there is an external drift acting on the particle \cite{WX15}. Another interesting future direction is to analytically study the tail distributions for self-avoiding RWs, which were investigated numerically in Refs.~\cite{Schawe2018,Schawe2019}. 
 Finally, it would be interesting to study the left tails for active particles (by analyzing their long-time survival probabilities in circles of given sizes) and also to study the $N\gg1$ limiting behaviors of the distributions \cite{Dewenter2016}, both for active and passive particles.




\bigskip

\subsection*{Acknowledgments}
We thank Baruch Meerson, Grégory Schehr and Satya Majumdar for useful discussions.
We are very grateful to Hendrik Schawe, Timo Dewenter and Alexander Hartmann for sending numerical data of Refs. \cite{Claussen2015, Dewenter2016, Schawe2020}. 
We acknowledge support from the Israel Science Foundation (ISF) through Grant No. 2651/23. SM thanks the LPTMS
for hospitality.   

\bigskip

\appendix
\section{ Minimal-action trajectories have constant speed} 
\label{OFM derivation}

In this appendix we show that the optimal paths have constant speed. We will show it for active particles with rate function $\Phi$, and treat BM as a particular  case in which $\Phi(z) \propto z^2$).
 For concreteness, we treat the case $d=3$, but the proof immediately extend to any dimension.




The action of a trajectory of an active particle in $d=3$, corresponding to the coarse-grained equation \eqref{LangevinCoarseGrained}, is
\be \label{actioneq}
s\left[\vec{r}\left(t\right)\right]=\int_0^T\Phi\left(\sqrt{\dot{x}^{2}+\dot{y}^{2}+\dot{z}^{2}}\right)\, dt \, .
\ee
Consider a given curve in the $xyz$ space, represented in parametric form by  
$(x_{0}\left(s\right), y_{0}\left(s\right), z_{0}\left(s\right))$
with $0 \le s \le 1$.
Then all time-dependent trajectories that follow this curve can be written in the form 
\be
(x\left(t\right), y\left(t\right), z\left(t\right)) = (x_{0}\left(f(t)\right), y_{0}\left(f(t)\right), z_{0}\left(f(t)\right))
\ee
for some monotonically-increasing function $f:[0,1]\to[0,T]$ that satisfies $f(0)=0$, $f(1)=T$.

Restricting ourselves to trajectories that follow such a curve, we rewrite the action \eqref{actioneq} as a functional of the function $f$. By using the chain rule in \eqref{actioneq} one obtains
\be
s\left[\vec{r}\left(t\right)\right]=\int_0^T\Phi\left( \dot{f} \sqrt{x_0'^{2}+y_0'^{2}+z_0'^{2}} \right)\, dt
\ee
where we use the shorthand notation $x_0' \equiv x_0'(f(t))$, and similarly for $y_0'$ and $z_0'$.
%
The Lagrangian 
 $L=\Phi\left( \dot{f} \sqrt{x_0'^{2}+y_0'^{2}+z_0'^{2}} \right)$
does not explicitly depend on time $t$, so the Hamiltonian $H$ is conserved. In order to calculate $H$, we calculate the conjugate momentum of $f$, 
\begin{equation}
    p = \frac{\partial L}{ \partial \dot{f}} = \Phi'\left(\dot{f} \sqrt{x_0'^{2}+y_0'^{2}+z_0'^{2}}\right)  \sqrt{x_0'^{2}+y_0'^{2}+z_0'^{2}} \, ,
\end{equation} 
which yields
\bea \label{hamiltonian}
H &=& \dot{f} p - L \nonumber \\
&=&  \dot{f} \sqrt{x_0'^{2}+y_0'^{2}+z_0'^{2}} \,\, \Phi'\left( \dot{f} \sqrt{x_0'^{2}+y_0'^{2}+z_0'^{2}}\right) \nonumber \\
&-& \Phi\left( \dot{f} \sqrt{x_0'^{2}+y_0'^{2}+z_0'^{2}}\right) \nonumber \\
&=& E = \,\,\, \text{constant in time.}
\eea
So $H$ is conserved in time.  One can rewrite $H$ in Eq. \eqref{hamiltonian} as a function only of the speed 
 $v=\sqrt{\dot{x}^{2}+\dot{y}^{2}+\dot{z}^{2}}$,
as
$H = v \, \Phi'(v) - \Phi(v)$.
It follows that the speed along the trajectory is constant in time, and is given by
\be
\sqrt{\dot{x}^{2}+\dot{y}^{2}+\dot{z}^{2}}=\text{const}=\mathcal{L}/T
\ee 
where $\mathcal{L}$ is the length of the curve 
$(x_0,y_{0}, z_0)$.
Plugging this back into Eq.~\eqref{actioneq} one finds that the action evaluated along the optimal trajectory (constrained on a given curve $(x_0,y_{0}, z_0)$) is given by
\be
\label{sOfTandL}
s\left[\vec{r}\left(t\right)\right]=\int_0^T\Phi\left(\frac{\mathcal{L}}{T}\right)\, dt = T \, \Phi\left(\frac{\mathcal{L}}{T}\right) \, .
\ee

The expression \eqref{sOfTandL} is an increasing function of $\mathcal{L}$, and therefore, its minimization with respect to curves $(x_0,y_{0}, z_0)$ (under various constraints such as convex-hull perimeter or area) boils down to the minimization of the length of the curve (under the constraints), as explained in the main text.

It is straightforward to extend this argument to arbitrary dimension $d>1$.






\begin{figure}[t]
\includegraphics[width=0.5\linewidth]{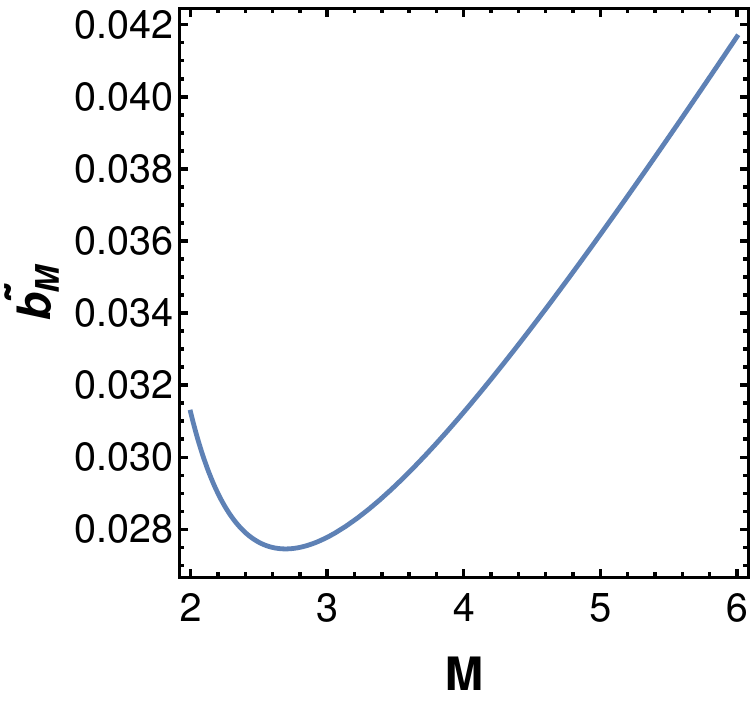}
\includegraphics[width=0.47\linewidth]{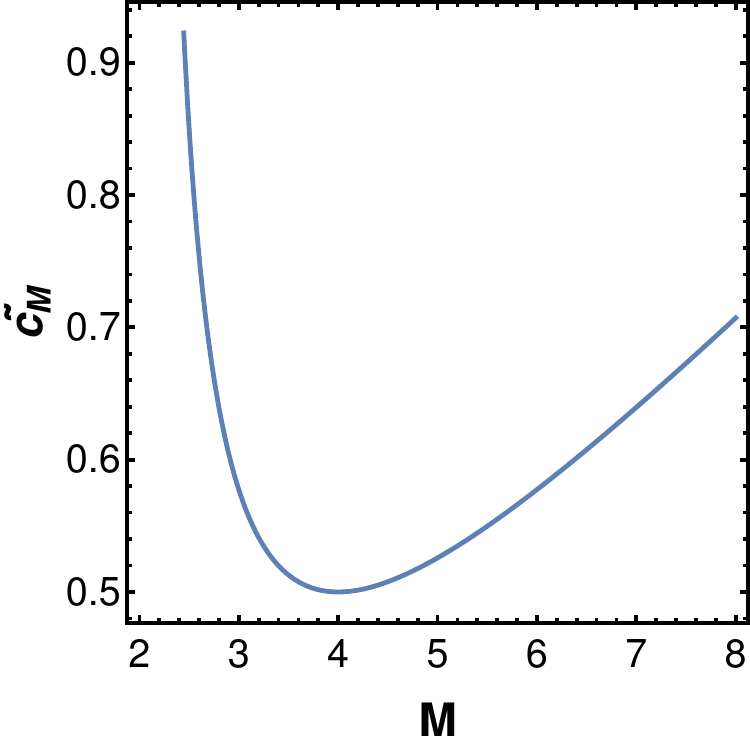}
\caption{ Coefficients $\tilde{b}_M$ and $\tilde{c}_M$ as a function of $M$. The  $M=3$ and $M=4$ are optimal solutions for perimeter and area respectively, i.e., they minimize the functions  $\tilde{b}_M$ and $\tilde{c}_M$ over integer values $M\ge 3$.}
\label{fig_bM-cM}
\end{figure}

\section{Multiple Brownian motions} \label{bmcm_plots}


         \begin{figure*}
\centering
     \begin{subfigure}[b]{0.4\textwidth}
     \centering
         \includegraphics[width=1\textwidth]{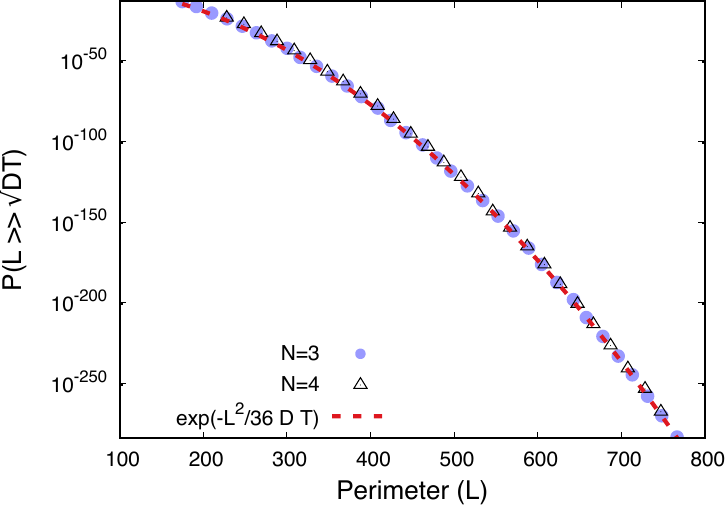}
         \caption{}
         \label{peri_mulBMcompare}
     \end{subfigure}
     \hfill
     \centering
     \begin{subfigure}[b]{0.4\textwidth}
     \centering
     \includegraphics[width=1\textwidth]{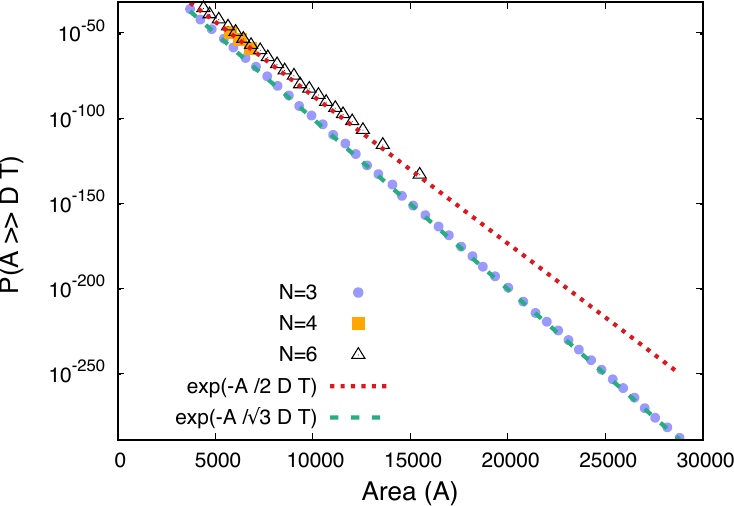}
         \caption{}
         \label{area_mulBMcompare}
     \end{subfigure}
     \hfill
        \caption{ 
        Right tail distribution of  \textbf{(a)} $\mathcal{P}(L)$ and \textbf{(b)} $\mathcal{P}(A)$ for multiple BMs. The markers depict the data of the distributions from \cite{Dewenter2016} and the dashed and dotted lines denote  our theoretical predictions  for the distribution tails. }
        \label{mulBMcompare}
\end{figure*}

\subsection{Showing that  $M=3,4$ are optimal} 

In this section, we explicitly show that $M=3$ and $M=4$ solutions are optimal for multiple BMsfor perimeter and area respectively, as stated in Sec. \ref{sec:multipleBMs}. As tabulated in Table \ref{tab:multiple rws}, the coefficients $\tilde{b}_M$ and $\tilde{c}_M$ of the action 
\be 
\tilde{b}_{M} = \frac{1}{16 M \sin^2{\frac{\pi}{M}}} \, , \qquad \tilde{c}_{M} = \frac{1}{2 \sin{\frac{2 \pi}{M}} } \, .
\ee
obtain their minima at $M=3$ and $M=4$ respectively.

 It is fairly easy to see that $\tilde{c}_M$ attains its minimum (for $M\ge2$) at $M=4$. This is because for $M=4$, the  sine function in the denominator attains its maximal value $1$.
Let us now show that $\tilde{b}_M$ attains its minimum (for integer $M > 2$) at $M=3$. For this, let us first minimize $\tilde{b}_M$ for real values of $M>2$, or equivalently, maximize
$g(M)= 1/\tilde{b}_M = 16 M \sin^2{\frac{\pi}{M}}$.
The requirement $g'(M)=0$ yields the transcendental equation
\be
\tan{\frac{ \pi}{M}} = {\frac{2 \pi}{M}} \, .
\ee
This equation has a unique (real) solution, $M=2.695\dots$, which corresponds to the global minimum of $\tilde{b}_M$ for real $M > 2$. Therefore, the minimum of $\tilde{b}_M$ for integer $M>2$ is at $M=3$.
 $\tilde{b}_M$ and $\tilde{c}_M$ are plotted, as functions of real $M$, in Fig.~\ref{fig_bM-cM}.

\subsection{Comparison with numerics}

In this subsection, we compare our expressions 
 \eqref{NBMsRightTailL} and \eqref{NBMsRightTailA}
for the right tails of the convex-hull perimeter and area distributions for $N$ non interacting BMswith the numerical data taken from \cite{Dewenter2016} (see Fig.~\ref{mulBMcompare}). We used \textit{webplotdigital} software to collect the data from the Figs. 11 and 12 of  Ref. \cite{Dewenter2016}. The numerical data provided in Ref. \cite{Dewenter2016} is for RW in discrete time and standard Gaussian step distribution.  The parameters that they chose were such that the time step was unity, corresponding to a diffusion coefficient $D = 1/2$, and they used $T = 50$. 

 For the perimeter distribution, we observe excellent agreement with our prediction for the coefficient $b_N = \frac
{1}{36}$, for all $N \geq 3$ (see Fig.~\ref{peri_mulBMcompare}). Indeed, we find that the numerical data for $N=3$ and $N=4$ both fall on the same theoretical curve.
For the area distribution, we find  that the numerical results are in excellent agreement with our predictions for the coefficients, $c_3 = 1/\sqrt{3}$ and
$c_N = 1/2$ for all $N \geq 4$.  Indeed, we find that the numerical results for $N=4$ and $N=6$ both fall on the same theoretical curve (see Fig.~\ref{area_mulBMcompare}).


\end{document}